\documentclass{article}
\usepackage{ijcai24}

\usepackage{times}
\usepackage{soul}
\usepackage{url}
\usepackage[hidelinks]{hyperref}
\usepackage[utf8]{inputenc}
\usepackage[small]{caption}
\usepackage{graphicx}
\usepackage{amsmath}
\usepackage{amsthm}
\usepackage{booktabs}
\usepackage{algorithm}
\usepackage{algorithmic}
\usepackage[switch]{lineno}
\usepackage{mathrsfs}
\usepackage{amssymb}
\usepackage{tabularx}

\usepackage{xcolor}

\title{Bridging Text and Molecule: A Survey on Multimodal Frameworks for Molecule}

\author{
Yi Xiao{\rm\textsuperscript{1}}\thanks{Equal contribution. $^\dagger$\,Corresponding author.}\and
Xiangxin Zhou{\rm\textsuperscript{1,2$\ast$}}\and
Qiang Liu{\rm\textsuperscript{1$\dagger$}}\And
Liang Wang{\rm\textsuperscript{1,2}}\\
\affiliations
\textsuperscript{1} Center for Research on Intelligent Perception and Computing (CRIPAC), \\State Key Laboratory of Multimodal Artificial Intelligence Systems (MAIS), \\Institute of Automation, Chinese Academy of Sciences (CASIA) \enspace \\
\textsuperscript{2} School of Artificial Intelligence, University of Chinese Academy of Sciences\enspace\\
\emails
\{y.xiao.cs\}@outlook.com,
\{zhouxiangxin1998\}@gmail.com,
\{qiang.liu,wangliang\}@nlpr.ia.ac.cn
}

\begin{document}

\maketitle

\begin{abstract}
Artificial intelligence has demonstrated immense potential in scientific research. Within molecular science, it is revolutionizing the traditional computer-aided paradigm, ushering in a new era of deep learning. With recent progress in multimodal learning and natural language processing, an emerging trend has targeted at building multimodal frameworks to jointly model molecules with textual domain knowledge. In this paper, we present the first systematic survey on multimodal frameworks for molecules research. Specifically,we begin with the development of molecular deep learning and point out the necessity to involve textual modality. Next, we focus on recent advances in text-molecule alignment methods, categorizing current models into two groups based on their architectures and listing relevant pre-training tasks. Furthermore, we delves into the utilization of large language models and prompting techniques for molecular tasks and present significant applications in drug discovery. Finally, we discuss the limitations in this field and highlight several promising directions for future research.
\end{abstract}

\section{Introduction}

Accurately modeling molecules and extracting meaningful features is a primary goal for molecular deep learning. Initially, manual descriptors such as molecular fingerprints and SMILES are proposed to describe molecules in strings or sequences. These descriptors can naturally be encoded by the transformer architecture for feature extraction. Subsequently, graph structures gradually shows their superiority in modeling the topology structure within molecules. Graph Neural Networks (GNNs) are employed to learn from molecular graph by aggregating and propagating information within atoms and chemical bonds \cite{kipf2017semi}. Simultaneously, numerous of works integrate self-supervised pre-training in this process to generate generalized representations. Despite the prosperous in molecular deep learning, two key challenges exist persistently. First, owing to the complexity of chemical space and chemical rules, current deep learning frameworks lack a deep comprehension of chemical domain knowledge (e.g. Quantum mechanics rules). Furthermore, both supervised and self-supervised models need to be trained or fine-tuned on labeled molecules, which are usually scarce in real application due to the costly experimental assessment. These notorious problems decelerate the progress in related areas.

Recently, multimodal learning and Large Language Models (LLMs) have shown impressive competence in modeling and inference. Inspired by the success of vision-language models, it is natural to associate molecules with text description to build multimodal frameworks. Following this idea, one line of works treat molecules as languages with special grammar, and cross-language frameworks, such as T5 \cite{raffel2020exploring}, are chosen as backbone to jointly model text and molecules. \cite{edwards2022translation,taylor2022galactica,pei2023biot5}. In the same time, another line of works explores the latent space alignment between text and structured molecular data \cite{su2022molecular,liu2023multi}, and attempts to integrate LLMs into multi-modal frameworks as predictor for cross-modal molecular tasks. Furthermore, prompting techniques are also introduced in the training process and yield competitive results in many molecular tasks without large-scale pre-training \cite{liang2023drugchat,cao2023instructmol,zhang2023moleculegpt}. Lately, some insightful works attempt to build autonomous agents for chemistry and biology \cite{boiko2023autonomous,liu2023chatgpt}, bringing new paradigm of future scientific research.

However, as a prosperous subject, there still lacks a systematic review to summarize recent progress and propose promising outlooks. In this regard, we present the first survey on multimodal frameworks for molecule. We summarize our contributions as follows: (1) We provide an overview of this field with a structured taxonomy that categorize the framework based on their basic architecture. (2) Our systematic review provides a detailed analysis of training strategies, dataset construction methods and corresponding applications. (3) We analyze the limitations in this field and provide several promising research directions.
\begin{figure*}[ht]
\begin{center}
\centerline{\includegraphics[width=\textwidth]{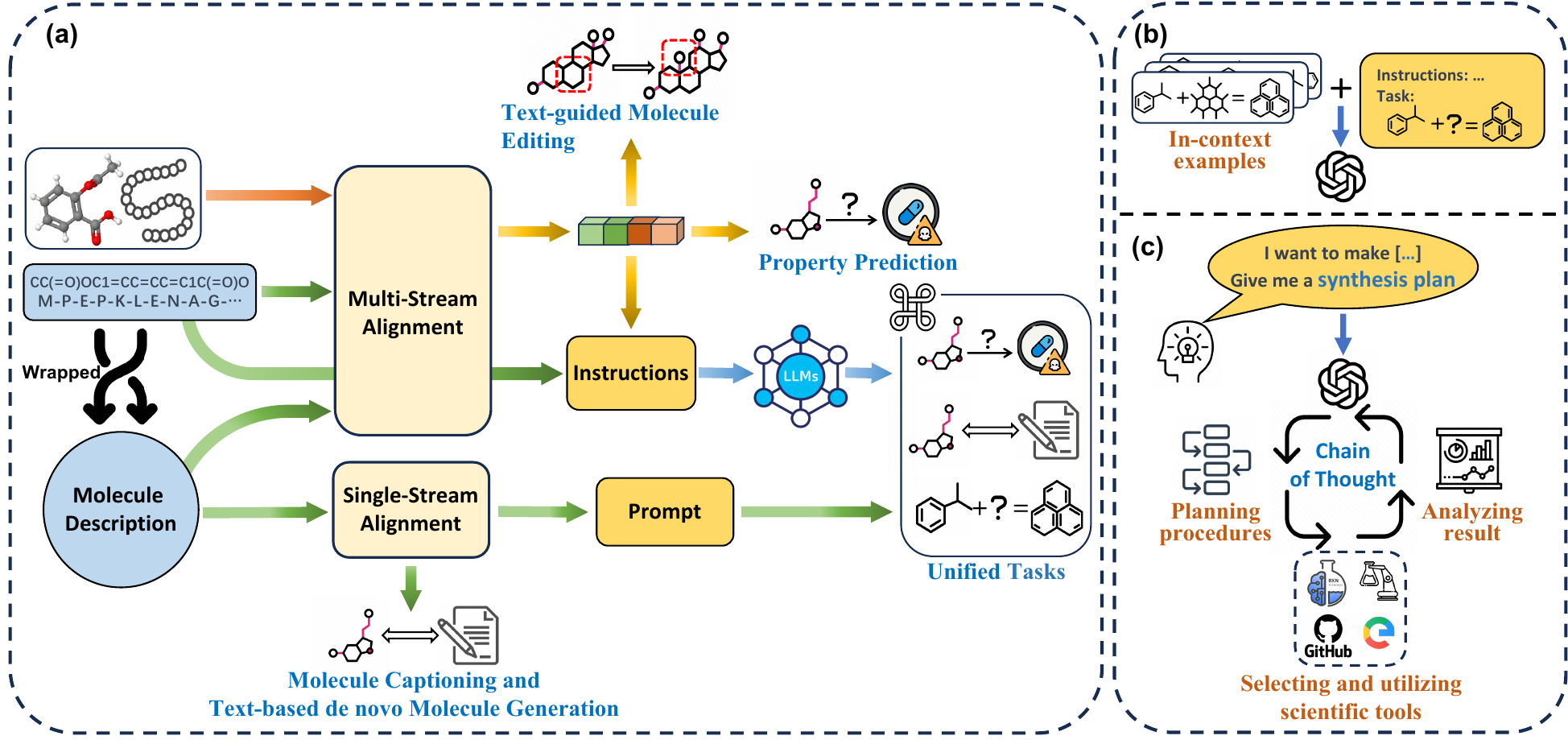}}
\vspace{-2mm}
\caption{Pipeline of multimodal framework for molecule and downstream molecular tasks (a-c). (a) Latent space alignment and adaptation of downstream tasks. The single-stream framework jointly models text and molecules with the same encoder. The downstream tasks are realized with task-specific prompts described in section \ref{prompt}; The multi-stream framework involves cross-modal alignment between text and molecules. Features from latent space can be directly used for tasks or be used in instruction-tuning. (b) Building a semi-autonomous agent for molecular research with instructions and in-context examples. (c) Building autonomous agent for chemistry with instructions and chain-of-thought prompting. Equipping agent with external tools and memory largely expand the autonomous level and capabilities.}
\label{fig:overview}
\vspace{-8mm}
\end{center}
\end{figure*}
\section{Molecular Descriptors and Encoding}
Molecules need to be transformed into descriptors for the recognition of the model. In this section, we briefly summarize the mainstream descriptors of small molecules and proteins, along with their corresponding encoder architectures. Generally, both small molecules and proteins can be described by sequences and graphs.
\subsection{Small Molecule Representation}
\paragraph{1D Sequences} The Simplified Molecular Input Line Entry System (SMILES) is the most frequently used sequential descriptor of small molecules, which maps atoms, bonds, and special structures with ASC\uppercase\expandafter{\romannumeral2} symbols. The molecular sequences can be tokenized like text sequence, and Transformer~\cite{vaswani2017attention} can be employed for molecular encoding \cite{zeng2022deep,liu-etal-2023-molxpt}. 
\paragraph{2D Graph} The topology structure of molecules can be naturally modeled by graph, with atoms as nodes and bonds as edges. GNNs~\cite{kipf2017semi} can be used to learn local and global representations of molecular graphs. Recently, the molecular representations that are pre-trained with GNNs structure have shown competitive results in various downstream tasks \cite{liu2022pretraining}, demonstrating the effectiveness of graph descriptors. 
\paragraph{3D Geometry} 2D molecular graphs have limitations in capturing the spatial information within molecules. For example, chiral molecules cannot be distinguished through most of the 2D graph. The geometry information of conformers (e.g., torsional angles, bond length) holds direct relation with molecular properties. In 3D geometry, atoms are associated with their coordinates with features expressed in high-order tensors to ensure geometric symmetries and expressiveness. Many studies concentrate on designing equivariant GNNs to accurately model the interaction between atoms \cite{batzner20223}.
\subsection{Protein Representation}
\paragraph{Protein Sequence} A protein can be viewed as a combination of 20 types of amino acids, making it possible to be expressed as amino acid sequences in a similar manner to molecules. The amino acid sequence captures the co-evolutionary information and plays a vital role in protein folding and function. Transformer-based models, commonly referred as protein language models (PLMs), use similar architecture in Natural Language Models to featurize protein for prediction or editing tasks.
\paragraph{Protein Graph} Protein functions are primarily determined by their folded structures \cite{jumper2021highly}. To better capture structural information, proteins can be represented as a residue-level relation graph, where nodes represent the alpha carbon of residues and edges represent connectivity between residues and amino-acids. We can also employ GNNs such as Massage Passing Neural Networks (MPNNs) for protein graph encoding \cite{dauparas2022robust}.

\section{Latent Space Alignment between Text and Molecule}\label{sec_align}
The encoding stage featurizes text and molecules into a single modality, while these representations still inhabit diverse semantic spaces and cannot interact with each other. To facilitate downstream tasks, different architectures are designed for text-molecule fusion and latent space alignment. In this section, we classify model architectures by the fusion scheme and summarize the corresponding pre-training tasks. We present a summary of representative works in Table \ref{tab:model}.

\subsection{Model Architecture}
Drawing inspiration from previous works in vision-language pre-training \cite{du2022survey}, we categorize models into \textit{single-stream}, and \textit{multi-stream} architecture. The two types of models mainly differ in their understanding of molecular latent space.
\paragraph{Single-Stream Architecture} A single-stream architecture assumes that the latent space of molecules and text share similar semantic meaning. Under this circumstance, molecules are treated as a specialized language and expressed by sequential descriptors such as SMILES. After tokenization, the molecular and textual tokens are typically fed into an encoder-decoder language model, such as T5~\cite{raffel2020exploring}, for multi-language pre-training. KV-PLM~\cite{zeng2022deep} and MolXPT~\cite{liu-etal-2023-molxpt} use byte-pair encoding (BPE) to tokenize SMILES and replace all molecular names in sequence with SMILES tokens, making these ``wrapped'' sequences as training data. BioT5~\cite{pei2023biot5} use separate vocabularies for molecules, proteins, and texts to avoid misunderstanding of tokens that may have the same expression but originate from different semantic spaces. Notably, GIMLET~\cite{zhao2023gimlet} serializes molecular graphs as node sequences and introduce position embedding to jointly encode nodes with related text tokens. This setting not only maintains the inductive bias at the graph level but also avoids introducing an extra graph encoding block and keeps molecular features independent of text.

\paragraph{Multi-Stream Architecture} Models with multi-stream architecture employ intra-modality processing for text and molecules. GNNs are generally adopted as molecular encoders to encode structural information. The representations are then projected into textual latent space by a linear layer, or projected with cross-attention mechanism for cross-modal fusion and alignment. For instance, \cite{abdine2023prot2text} fuse protein sequence and protein graph features by element-wise addition and use them as input of cross-attention module to adapt with text. \cite{xu2023protst} choose both text and protein representation as keys and apply two separate cross-attention modules to produce fused-text and fused-protein representations.

Q-Former~\cite{li2023blip} is a representative architecture in vision-language modeling that leverages cross-attention layers to bridge the modality gap. Similarly, \cite{li20243d,liu2023molca,zhang2023moleculegpt} adopt Q-Former to connect molecular graph with text and extract text-related molecular features with a learnable query. Particularly, \cite{liu2023git} propose GIT-Former which can be viewed as a variation of Q-Former with additional input modalities from molecular images and sequences. 

\subsection{Pre-training Tasks}
The fused representations need to be aligned in a unified latent space to keep consistent semantic meaning for downstream tasks. In this section, we review the commonly used pre-training tasks for alignment between text and molecules.   
\paragraph{Molecule-Text Contrastive Learning} The contrastive learning (CL) task between molecule and text aims to align multi-modal representations by enhancing the correlation between matched molecule-text pairs. The contrastive learning objective pushes the embeddings from matched text and molecules closer in latent space while enlarging the distance between pairs from different molecules. The CL task will enhances the model with cross-modal retrieval and matching ability. Here, we present the expression of commonly used InfoNCE~\cite{oord2018representation} loss:
\begin{align}
    \mathcal{L}_{\text{NCE}} = -\sum_{i} \log\frac{\exp(z_i^M\cdot z_i^T / \tau)}{\sum_{j=1}^N \exp(z_i^M \cdot z_j^T / \tau)}
\end{align}
where $\tau$ is the temperature coefficient. In order to facilitate the convergence, a trainable linear projector can be used to minimize the modality gap before the contrastive learning \cite{liu2023multi}. \\
\indent Although contrastive learning is an effective approach for cross-modal molecule-text alignment, it has some domain-specific limitations. One issue is that the structural information of molecules will lost in the process of encoding. Additionally, limited number of molecule-text pairs also has impacts on the alignment result. Motivated by the molecular graph augmentation methods \cite{you2020graph}, we could adopt augmented graphs to construct molecule-text pairs, as the augmented graphs remain similar semantics information to the original graphs, such as properties and structures. Applying a contrastive loss between the embeddings of the original graph and the augmented graph also ensures this consistency. MoMu~\cite{su2022molecular} introduces two augmented graphs with node dropping and sub-graph extraction to construct matched pairs with text descriptions. MolLM~\cite{luo2023unified} follows the same augmentation rules in MoMu and introduces two extra augmentations which are chemical transformation and motif removal. These augmentations increase the size of training data, making the alignment process more robust.
\paragraph{Molecule-Text Matching} Molecule-text matching (MTM) aims to predict whether a molecule-text pair is matched or not. It is defined as a binary classification task with the following loss function:
\begin{align}
    \mathcal{L}_{\text{MTM}} = -\mathbb{E}\left[ \sum_{i} [\log p(m_i,t_i) -\right.\notag \\
    \left.\log p(m_i,t_j) - \log p(m_j,y_i)] \right]
\end{align}
where $(m_i,t_i)$ denote matched molecule and text pair. The MTM task enables the model with the retrieval ability and refines the alignment between text and molecule. \cite{liu2023molca,li20243d} use MTM in the first-stage training of Q-Former. \cite{liu2023git} extend the MTM to cross-modal matching which performs matching tasks between text, graphs, and images.

\paragraph{Conditional Generation} Conditional generation (CG) aims to generate tokens based on given conditions or constraints. Tasks such as molecule captioning and text-based molecule generation all fall into this category. Conditional generation enables models to learn complex mapping rules between text and molecules. It is adaptable for T5 architecture where all molecular tasks are transformed into text-to-text generation format. The objective function can be written as:
\begin{align}
    \mathcal{L}_{\text{CG}} = - \sum _{\substack{i}}^{\substack{n_i}} \log P(u_i|C;\theta)
\end{align}
where $u_i$ is the $i$-th token and $C$ denotes the generation condition which may be referred to as a molecule graph or text description depending on the task.

\paragraph{Masked Language Modeling} As discussed in section \ref{sec_align}, modeling languages and molecules may share similarities. Under this assumption, masked language modeling as a prevalent pre-training task for LLMs can also be used for training molecule sequences or wrapped sequences. During the pre-training stage, the models are trained to predict the masked components using the remaining context. The training objective is defined by cross-entropy
\begin{align}
    \mathcal{L}_{\text{MLM}} = -\mathbb{E}_{T\in \mathcal{D}} \sum_{\substack{\Tilde{m}\in \mathcal{M}}}\log p(\Tilde{m}|T\backslash \mathcal{M})
\end{align}
where $\mathcal{M},T\backslash \mathcal{M},T$ represent the masked tokens, unmasked tokens, tokenized text and molecules separately. This self-supervised pre-training task can enhance the contextual comprehension of the model, improving performance in many downstream tasks. For MLM, there are two types of masking: \textbf{token masking} represented by BERT~\cite{devlin2018bert} and its variants, and \textbf{span masking} represented by T5~\cite{raffel2020exploring}. According to \cite{raffel2020exploring}, span masking is more efficient. \cite{zeng2022deep} randomly mask tokens from both molecules and text in wrapped text sequences. \cite{edwards2022translation,pei2023biot5,rubungo2023llm,qian2023predictive} all adopt span masking for MLM with corresponding T5 backbone to enhance the downstream translation tasks between molecule and text. \\
\indent Similarly, the protein language models (PLM) introduce masked protein modeling (MPM) by masking residues in protein sequences. ProtST~\cite{xu2023protst} not only uses MPM for protein encoder pre-training, it also utilizes fused text and protein representations to predict the type of masked residues and language tokens. Before entering the PLM and biomedical language model, 15\% of residues in protein sequence and 15\% of word tokens in text are masked and then embedded. Two MLPs are trained with MPM objective to recover the masked components in text and residues.
\paragraph{Casual Language Modeling} Different from the autoencoder (AE) language models such as BERT and T5 which adopt MLM as the training objective, the autoregressive models represented by GPT~\cite{yenduri2023generative} are trained with Casual Language Modeling (CLM). The objective of CLM is to predict the next token in a sequence in a left-to-right direction. The objective function can be written as
\begin{align}
    \mathcal{L}_{\text{CLM}} = - \sum _{\substack{i}} ^{\substack{n_i}} \log P(u_i|u_{i-k},...,u_{i-1};\theta)
\end{align}
where $n_i$ and $k$ represent the number of tokens and context length. Transformation of molecular tasks into text generation helps the CLM seamlessly integrate into the training and instruction-tuning process. \cite{liang2023drugchat,cao2023instructmol,zhang2023moleculegpt}. We will discuss the detail of instruction-tuning and adaptation of tasks in the following sections.

\section{Bridging LLMs and Molecular Tasks with Prompting Techniques}
With the advancement of multimodal large language model (MLLM), the cross-modal inference ability of LLMs could be extended to chemistry research. Compared with traditional cross-modal learning which focuses on modality alignment, MLLM leverage powerful LLM as the brain to process multi-modal information and utilize multiple prompting techniques such as instruction-tuning (IT), in-context learning (ICL) and chain-of-thought (CoT) to bridge LLMs with downstream tasks \cite{li2023blip}. As shown in Figure \ref{fig:overview}, LLMs could conduct multiple molecular tasks with instructions and cross-modal input. In this section, we discuss the prompting techniques to build MLLM in molecular science, and show the progress to build intelligent agents for chemistry.  
\subsection{Prompt-based Fine-tuning on LLM} \label{prompt}

To bridge the gap between pre-training and downstream tasks, \cite{raffel2020exploring} transfer all NLP tasks into text-to-text generation format with task-specific prefix. Based on this work, \cite{gao-etal-2021-making} propose prompt-based fine-tuning to unify the fine-tuning framework among different tasks with task-specific prompts. This strategy can also be applied to unify different molecular tasks for better adaptation in single framework. For example, the prompt of BBBP property prediction task in MoleculeNet~\cite{wu2018moleculenet} can be designed as: ``\emph{We can conclude that the BBBP of \textless SMILES\textgreater{} is \textless tag{}\textgreater}" where \textless tag\textgreater{} is the ``true'' or ``false'' prediction given by model \cite{liu-etal-2023-molxpt}. In this way, we unify the prediction task into a text generation format. Then model is fine-tuned and evaluated on each task with pre-training parameters fixed. \cite{pei2023biot5} enrich the above-mentioned template with task definition and explanation, which brings improvement in property prediction accuracy. \cite{liu2023molca} integrate fused feature as soft prompt and use LoRA~\cite{hu2021lora} to improve adaptation efficiency. Compared with traditional fine-tuning, prompt-based fine-tuning shows impressive performance in few-shot datasets.

\subsection{Instruction Tuning on LLM for Zero-shot Learning Ability}
Unlike prompt-based tuning, instruction tuning \cite{wei2022finetuned} aims to adapt models to various tasks. In the tuning process, models are trained in multiple tasks which have been unified through task-specific instructions. This multi-task learning strategy enables models to comprehend instructions and to seamlessly transfer to various tasks in a few-shot or zero-shot manner. A standard instruction entry is typically composed of three main parts: an \emph{\textless instruction\textgreater{}} that clarifies the task, an \emph{\textless input\textgreater{}} which is usually molecular feature, and an \emph{\textless output\textgreater{}} that embodies the expected outcome \cite{fang2023mol}. 
\cite{liang2023drugchat,luo2023biomedgpt,cao2023instructmol,li20243d,zhang2023moleculegpt} use the fused feature as a soft prompt to compose the instructions. During the tuning process, the fusion architecture is fine-tuned solely and some works also use LoRA~\cite{hu2021lora} to improve efficiency \cite{li20243d,cao2023instructmol}. 
\cite{zhao2023gimlet} compared the instruction-tuned GIMLET with other pre-trained baseline in zero-shot property prediction tasks. The leading accuracy shows the strong generalization performance to novel tasks by following instructions.
\subsection{In-Context Learning and Chain-of-Thought} \label{ICL}

Recently, various of attempts have been made to integrate LLMs into various chemistry research as an intelligent agent, with wide range of applications such as autonomous experiment planning \cite{bran2023augmenting,boiko2023autonomous}, conversational drug editing \cite{liu2023chatgpt}, chemical reaction prediction \cite{shi2023relm}, etc. These models leverage the in-context learning (ICL) or chain-of-thought (CoT) prompting~\cite{wei2022chain} which enables step by step reasoning and few-shot prediction for specific tasks. 
The in-context learning for molecular tasks usually combines instruction-based prompts with a few molecular Question-Answer examples.
\cite{chen2023artificially,li2023empowering} design the few-shot prompting with role definition, task description, in-context examples and output control to guide the prediction of LLM. Differently, ReLM~\cite{shi2023relm} enhances the reaction prediction result from a GNN-based model by integrating LLM as a decision-maker. LLM learns to self-evaluate its prediction from in-context examples with confidence scores.

The autonomous reasoning of LLM agents can be achieved by chain-of thought prompting with few-shot or zero-shot manner. The few-shot CoT directly demonstrates the reasoning steps in one or few prompts, and the agent can leverage the emergent ability of LLMs to imitate similar reasoning in the same type of tasks. Moreover, the zero-shot CoT simplifies the prompt with ``Let's think step by step'' at the end of the problem description. With effective CoT and access to external knowledge, LLM agents can work semi-autonomously to support experts in scientific research. In STRUCTCHEM~\cite{ouyang2023structured}, GPT-4 is guided to solve chemistry problems through formula generation and step-by-step reasoning. To correct the errors in CoT reasoning, another GPT-4 is employed to perform iterative review-and-refinement for generated results in each step. ChemCrow~\cite{bran2023augmenting} adopt Least to Most prompting~\cite{zhou2023leasttomost} (LtM) which can be seen as CoT in an auto-regressive manner. The reasoning loop in ChemCrow integrates the decomposition of the task, selecting and using external tools, and analysing the result. The input of the next reasoning loop is built upon the current results until they satisfy the expected format. It is the first LLM agent in chemistry capable of automatically completing complex planning and synthesis task.

\section{Dataset Construction}
\begin{table*}[t]
    \centering
    {\fontsize{7}{7}\selectfont
    \setlength{\tabcolsep}{3pt}
    \begin{tabular}{llllll}
        \toprule
        \textbf{Model} & \textbf{Molecule descriptors}  & \textbf{Backbone architecture}        & \textbf{Training database} & \textbf{Training task}  \\
        \midrule
        MolT5 ~\cite{edwards2022translation} & SMILES  & T5 & C4 + ZINC & MLM \\
        Galactica ~\cite{taylor2022galactica} & Bio-Sequence   & Transformer Decoder & Not open  & CLM \\
        KV-PLM ~\cite{zeng2022deep} & SMILES & SciBERT~\cite{beltagy-etal-2019-scibert} & PubChem + S2orc & MLM \\
        MolXPT~\cite{liu-etal-2023-molxpt}	& SMILES	& GPT &	PubMed + PubChem	& CLM \\
        BioT5~\cite{pei2023biot5} &	SELFIES + Protein Sequence	& T5 & PubChem + Swiss-Prot & MLM + CG\\
        Text + Chem T5~\cite{christofidellis2023unifying}	& SMILES	& T5	& Multi-domain & CG    \\
        TextReact~\cite{qian2023predictive} &SMILES  &  SciBERT &	USPTO	&	CL + MLM + CG   \\
        GIMLET~\cite{zhao2023gimlet} & Graph &T5	& ChEMBL 	& CG	\\  
        LLM-Prop~\cite{rubungo2023llm}	& Molecule Strings & T5 & Materials Project	 & MLM   \\
        \midrule
        Text2Mol~\cite{edwards2021text2mol}  & Graph & Multi-stream + Transformer & ChEBI-20 & CL \\
        MoMu~\cite{su2022molecular}& Graph & Multi-stream& PubChem + S2orc & CL \\
        MolLM~\cite{tang2023mollm} & SMILES + Graph + Geometry & Multi-stream & PubChem + S2orc &CL \\
        DrugChat~\cite{liang2023drugchat} &  Graph  & Multi-stream + Vicuna-13b	& PubChem & CLM	\\
        MoleculeSTM~\cite{liu2023multi} & Graph	& Multi-stream + Decoder	&PubChem	& CL	\\	
        BioMedGPT~\cite{luo2023biomedgpt}& Graph + Protein Sequence & Multi-stream + LLaMA 2 & PubChem + S2orc + UniProt & CLM \\
        InstructMol~\cite{cao2023instructmol} & SELFIES + Graph & Multi-stream + Vicuna-7b & PubChem &CLM \\

        GIT-Mol~\cite{liu2023git} &	SMILES + Graph + Image	& Q-Former + T5	&PubChem &	MTM + CL \\

	CLAMP~\cite{seidl2023enhancing} & Fingerprints & Multi-stream & PubChem & CL \\					
        MolCA~\cite{liu2023molca} &	SMILES + Graph &	Q-Former + Llama 2	& PubChem	&MTM + CL + MC + CLM		\\						

        3D-MoLM~\cite{li20243d} &	SMILES + Geometry	& Q-Former + Llama 2	& PubChem		& MTM + CL + MC + CLM \\
        MoleculeGPT~\cite{zhang2023moleculegpt}& SMILES + Graph	& Q-Former + Vicuna-7b &	PubChem		&CL+CLM \\

        ProtST~\cite{xu2023protst}	& Protein Sequence	& Multi-stream	& SwissProt		& CL + MLM	\\
        ProtDT~\cite{liu2023text} & Protein Sequence  & Multi-stream + Decoder &	SwissProt	& CL  \\
        Prot2Text~\cite{abdine2023prot2text} & 	Protein Sequence + Protein Graph 	& Multi-stream + Transformer &	SwissProt &	CLM	  \\
        InstructProtein~\cite{wang2023instructprotein} & Protein Sequence & Knowledge Graph + LLMs & UniProt & CLM\\
        BioBridge~\cite{wang2024biobridge} & SMILES + Protein Sequence & Knowledge & PrimeKG etc. & CL \\
        \midrule
        ReLM~\cite{shi2023relm} & SMILES + IUPAC + Graph & ICL + LLMs & - & - \\
        ChemCrow~\cite{bran2023augmenting} & - & CoT + LLMs & - & - \\
        ChatDrug~\cite{liu2023chatgpt} & SMILES & LLMs & - & - \\
        MolReGPT~\cite{li2023empowering} & SMILES & ICL + GPT-3.5 & - & - \\
        \bottomrule
    \end{tabular}
    }
    \caption{Summary of representative multimodal frameworks}
    \label{tab:model}
\end{table*}

The quality of the training data is crucial for modal alignment and training, significantly influencing the performance of the model. In this section, we focus on dataset construction methods. The data resource can be found in Table \ref{tab:model}.

\paragraph{Data Processing}
To facilitate modality alignment, pairs of textual and non-textual molecular data are collected from public datasets. However, the descriptions in databases are not balanced. Taking PubChem~\cite{kim2023pubchem} as an example, it is very often that some molecules only have a few basic records and lack some corresponding properties. To tackle this issue, many researchers construct training data from multiple datasets or retrieve relevant text from scientific corpus such as S2orc~\cite{lo2019s2orc}. Meanwhile, the pre-processing methods are also important. For instance,  \cite{liu2023multi,zhang2023moleculegpt,cao2023instructmol} first replace all the molecule names in the annotation of PubChem with token `$\sim$' to simplify name comprehension in training. Then they remove the redundant information in the molecule description such as origins, sources, and some geographic notations that have no relation with structure or property. \cite{xu2023protst} select four kinds of properties from  Swiss-Prot~\cite{bairoch2000swiss} and use fixed templates to rearrange the descriptions, ensuring the consistency of training data format.
\paragraph{Integrating Generative AI}
Recent advances in generative AI provide an innovative approach to mitigate the data scarcity challenge. For instance, \cite{li20243d}
use GPT-3.5 to enrich the sparse molecular descriptions in PubChem. \cite{fang2023mol} leverage GPT-3.5 to diversify prompt templates and use them to generate QA pairs for instruction-tuning. Additionally, \cite{sakhinana2023crossing} use GPT-4 to generate molecule captions for the fine-tuning. \cite{chen2023artificially} fabricate an ``artificially-real'' dataset for domain adaptation, where molecule descriptions are generated through ChatGPT with retrieval-based few-shot prompting.

\section{Applications}
This section will showcase applications in drug discovery and chemistry research employing the aforementioned methods. Beyond the introduction of tasks, our emphasis lies in the adaptation between the model and tasks.
\subsection{Text-molecule Retrieval}
Text-molecule retrieval task is first proposed by \cite{edwards2021text2mol}, which aims to retrieve the corresponding molecule from a given text query. This molecule retrieval task is useful in early stages of drug discovery, where experts need to select potential molecules from compounds database for further design and optimization. The retrieval task can be accomplished by the aligned latent space between language and molecules, from which we can acquire the encoded text descriptions with the connection of target molecules. Then we can use the similarity score to evaluate the distance between text and molecules to find the best-matched pair. In KV-PLM~\cite{zeng2022deep}, descriptions and molecules are encoded by a shared transformer encoder. MoMu~\cite{su2022molecular} and MoleculeSTM~\cite{liu2023multi} use separate encoders to extract multimodal features and align the latent space with contrastive learning.
\subsection{Property Prediction}  
One of the important goals of drug discovery is to search for small molecules and proteins with desired properties. The molecule description from scientific literature and databases serve as knowledge repositories that contain properties, interactions, and structures that can hardly be inferred from current models \cite{pei2023biot5}. Through molecule-text alignment, text information can act as a supplementary signal to enhance molecular representation and improve the performance of models in property prediction \cite{seidl2023enhancing,xu2023protst}. 
The property prediction task is usually a binary classification task achieved by molecular features and simple prediction head. An alternative approach is to leverage powerful generative LLMs with instructions to predict property in a QA format \cite{zhang2023moleculegpt,liu2023git}. As shown in \ref{prompt}, the prediction is determined by the probabilities of tokens ``true'' and ``false'' in the generated answer. 
\subsection{Molecule Design}
\paragraph{\textit{De novo} Generation}
\textit{De novo} generation in molecule design includes molecule captioning which generates a description of given molecules and text-guided \textit{de novo} generation which generates molecules from scratch by the textual guidance. Models with single-stream architecture have the privilege of performing translation between text and molecule, owing to the encoder-decoder structure and text-to-text task format. \cite{raffel2020exploring}. Apart from the translation-based methods. \cite{liu2023text} propose a protein design framework with a multi-stream encoder. In text-guided protein generation task, the description is first encoded by the aligned text encoder. Then a facilitator module which is parameterized by a multi-layer perception is used to learn the transformation from encoded text to protein representation. The resulting protein representation is then fed into a trained generative decoder to generate protein sequences.
\paragraph{Molecule Editing}
Molecule editing seeks to optimize current molecules with desired properties. Within the drug discovery pipeline, text-guided editing finds application in lead optimization tasks and proves valuable for decomposing multi-objective lead optimization \cite{liu2023text}. Drawing inspiration from the success of few-shot text-to-image generation, text description can simplify the complexity of the target chemical space in the generation process. Simultaneously, personalized generation enhances drug editing by introducing high flexibility.
As mentioned above, the latent space alignment establishes a unified latent space where features possess semantic meaning in both structure and text. Building upon this approach, \cite{liu2023multi,liu2023text,tang2023mollm} use latent optimization methods to sample a latent representation close to both text and molecule in latent space. Then this latent code is fed into a decoder which is usually a trained molecule generation model to produce optimized molecules. \cite{kim2023dataefficient} proposes hierarchical textual inversion which introduces intermediate and detail tokens to represent SMILES, aiming to capture cluster-level and molecule-level features. The interpolation sampling can benefit from this hierarchical design with high generation diversity.

\subsection{Other Applications}
\paragraph{Reaction Prediction} Reaction prediction is a challenging but fundamental task in chemistry and biology. The chemical reaction process can be seen as a mapping between a set of reactants to a set of products with specific reaction conditions. Based on this assumption, there are three main tasks in reaction prediction, which are product prediction, reaction condition prediction, and most importantly, retrosynthesis prediction. Text can supply information in complex reaction mechanisms and reaction templates which GNN-based methods often fail to capture. \cite{qian2023predictive} retrieve reaction-related text and concatenate with input similes to enhance the retrosynthesis prediction. As described in \ref{ICL}, we can also involve LLMs in reaction prediction via prompting engineering. For example, \cite{shi2023relm} use GPT-4 to predict reaction products with the aid of in-context reaction examples and reaction prediction model. 

\paragraph{Intelligent Agent for Scientific Research}
According to \cite{bran2023augmenting}, the automation level in chemistry is relatively low compared to other domains. Despite LLMs may have difficulties in comprehending chemistry principles, they have demonstrated significant capability in understanding human instructions and organizing information based on extensive training corpora \cite{ai4science2023impact}. Consequently, LLMs have the potential to become intelligent assistants to automatically arrange research with the help of professional tools and software. \cite{liu2023chatgpt} design a drug editing agent with conversational interaction. The agent can receive human feedback to retrieve candidate drug molecules from the database with desired properties. \cite{boiko2023autonomous} develop a ``Coscientist'' based on GPT-4 similar to ChemCrow~\cite{bran2023augmenting} which can autonomously design and execute chemical research.

\section{Conclusions and Future Outlooks}
In this paper, we provide a comprehensive review of multimodal frameworks for molecules. After a brief introduction to the background and molecule descriptors, we introduce the model architecture and pre-training tasks for latent space alignment. Then, we summarize the prompting techniques in a multimodal large language model to bridge LLMs with molecular tasks. As an application-oriented domain, we combine the aforementioned methods to exhibit applications in drug discovery. Although text-molecule models have made impressive progress, there exist several challenges which appeal to future research. 
\subsection{Appealing for High-Quality Data and Reliable Benchmarks}
According to the neural scaling law, the emergent abilities of LLM in complex molecular tasks have not been shown. The data scarcity challenge still exists for both molecules and text description data. In addition to collecting descriptions from databases, many works also automatically retrieve relative text from scientific corpus, while the authenticity and correlation of retrieved text cannot be guaranteed \cite{xu2023protst,tang2023mollm}. For the progress of the community, a larger and more qualified molecule-text database is significant.
Although multimodal frameworks exhibit great potential in various molecular tasks, there remains a question of how to fairly evaluate the performance among different models. Experimental results may be unreliable due to inconsistent settings between different models and low representative of test datasets. To address this concern, new benchmarks are necessary to standardize evaluation metrics and settings, providing more reliable and realistic test data, such as drug-like molecular datasets. Several attempts have been made in this direction \cite{guo2023large}.

\subsection{Extending the Interpretability of Model}
The lack of interpretability prohibits many applications of deep molecular models as numerical predictions alone may not be convincing enough compared with computational and experimental results. Text-involved multimodal frameworks provide an opportunity to enhance the interpretability of results. By leveraging in-context learning and chain-of-thought prompting in LLMs, models can reasoning and inference, like the human brain, to produce explainable results. Follow-up research can also try to develop interpretable tools to bridge the relation between textural description and molecule structure in latent space \cite{su2022molecular}. Furthermore, \cite{wellawatte2023extracting} explore the possibility of combining XAI methods with LLMs to provide explanations of the structure-property relationship in a comprehensive way.

\subsection{Improving the Reasoning Ability}
Applying prompting techniques can significantly improve the reasoning ability of LLM-based frameworks. However, it is observed that in some cases, models may generate unrealistic predictions or even replicate the values in examples as prediction \cite{zhao2023what}. This serves as an evidence that LLMs may rely on memorization without truly understanding the molecules and chemical problems. Future studies may integrate successful GNNs into transformer-based model architecture, other than simply using GNNs as encoders \cite{zachares2023form}. Designing effective prompts for molecular tasks can also be taken into consideration.

\subsection{Integration with Foundation Models}
Foundation models (FMs) in the biomedical domain have shown promising performance. For example, AlphaFold~\cite{jumper2021highly} can accurately predict protein's structure from amino-acid sequence. These foundation models are usually uni-modal with sufficient training data. It is possible to integrate FMs within LLM agents or specially designed frameworks. \cite{wang2024biobridge} has tried to model the relation between the FMs and the knowledge graph. We believe that effective frameworks could unlock the additive power of FMs.

\subsection{Learning from Human/AI Feedback} Recent progress in reinforcement learning from human/AI feedback (i.e., RLHF~\cite{ouyang2022training} and RLAIF~\cite{lee2023rlaif}) has achieved promising results in aligning LLMs with human preference. RLHF fits a reward model to human preference dataset and use RL to optimize LLMs to produce responses assigned with high rewards. This paradigm may pave the way for utilizing LLMs for biomedical applications, especially in scenarios where molecular simulation software can be used as a reward model. Exploring how to fully utilize the power of RLHF at the interaction of text and molecules is an appealing research direction.

\clearpage
\bibliographystyle{named}
{\footnotesize
\bibliography{main}}

\end{document}